\documentclass[11pt,a4paper, twocolumn]{article}

\usepackage[utf8]{inputenc}
\usepackage[T1]{fontenc}
\usepackage[english]{babel}

\usepackage{geometry}
\usepackage{amsmath, amssymb, amsthm}

\usepackage{graphicx}
\usepackage{caption}
\usepackage{subcaption}

\usepackage{academicons}
\usepackage{scalerel} % pour aligner l’icône correctement

\usepackage[backend=biber,sorting=none]{biblatex}
\AtEveryBibitem{%
  \clearfield{urldate}%
}

\usepackage{xcolor}
\usepackage{hyperref}
\hypersetup{
    colorlinks=true,
    linkcolor=gray,
    citecolor=gray,
    urlcolor=gray
}

\title{Modeling Interactive Narrative Systems: A Formal Approach}
\author{
    Jules Clerc\textsuperscript{1,2}~(\href{https://orcid.org/0009-0009-0825-4498}{ORCID}),
    Domitile Lourdeaux\textsuperscript{1}~(\href{https://orcid.org/0000-0002-3354-7294}{ORCID}),
    Mohamed Sallak\textsuperscript{1}~(\href{https://orcid.org/0000-0003-4803-2362}{ORCID}),\\
    Johann Barbier\textsuperscript{3},
    Marc Ravaine\textsuperscript{2}
}

\date{
    \textsuperscript{1}Université de technologie de Compiègne, CNRS, Alliance Sorbonne Université, Heudiasyc, Compiègne, France\\
    \textsuperscript{2}ALTEN, Sophia Antipolis, France\\
    \textsuperscript{3}Safe Innovation, France\\[1ex]
}

\begin{document}

\maketitle

\renewcommand{\thefootnote}{}
\footnotetext{
  \textbf{Contact:} \\
  jules.clerc@hds.utc.fr, \\
  domitile.lourdeaux@hds.utc.fr, \\
  mohamed.sallak@hds.utc.fr, \\
  johann.barbier@safe-innovation.com, \\
  marc.ravaine@alten.com
}
\renewcommand{\thefootnote}{\arabic{footnote}}

\begin{abstract}
  Interactive Narrative Systems (INS) have revolutionized digital experiences by empowering users to actively shape their stories, diverging from traditional passive storytelling. However, the field faces challenges due to fragmented research efforts and diverse system representations. This paper introduces a formal representation framework for INS, inspired by diverse approaches from the state of the art. By providing a consistent vocabulary and modeling structure, the framework facilitates the analysis, the description and comparison of INS properties. Experimental validations on the "Little Red Riding Hood" scenario highlight the usefulness of the proposed formalism and its impact on improving the evaluation of INS. This work aims to foster collaboration and coherence within the INS research community by proposing a methodology for formally representing these systems.
\end{abstract}

\smallskip


\section{Introduction}
Unlike traditional narratives where viewers are spectators, INS allow users to directly influence the course of a story. These systems find applications in various sectors, including entertainment, education and training, with prominent examples such as “Choose Your Own Adventure” books and interactive systems like Façade \cite{mateasFacadeExperimentBuilding2003a}. The field has recently seen new research activity since the arrival of LLM, that excel in a wide range of tasks, have enabled many advances in the field of artificial intelligence, particularly concerning text generation and natural language understanding \cite{naveedComprehensiveOverviewLarge2023}. However, despite the arrival of these new revolutionary tools, the field of interactive narrations still suffers from certain problems such as the absence of a formalism commonly used by the entire community. This problem is even more important today, because of the hype caused by these AI, which is leading to the creation of many new INS that are difficult to analyse and compare due to a lack of scientific framework \cite{koenitzCanAICreate2025}.

To address these challenges, this work proposes a methodology which aims to model and study the properties of INS. By developing systems around a coherent structure, we aim to eliminate ambiguities and provide a basis for future work, such as the possibility of defining and comparing INS properties from a functional analysis. This approach is part of a broader effort by the research community to coordinate and pool knowledge, as highlighted by the Association for Research in Digital Interactive Narratives (ARDIN) \cite{ardin_online}.

In the following sections, we detail the proposed formalization, illustrate its application through experiments, and explore its implications for the evaluation and development of INS.

\section{Related Work}
The lack of consensus extends to terminology, which varies across application contexts such as gaming, training, or education. To ensure clarity and consistency, we propose a set of definitions.
\begin{itemize}
    \item \textit{Player}: The individual who engages with the system and performs actions. In some literature, this role is referred to as the "user" or the "trainee."
    \item \textit{Creator}: The individual responsible for designing the system, including its states, transitions, and overall structure. Other terms for this role include "developer" or "author."
    \item \textit{Experience Manager (EM)}: A system-side "player" who has complete control and knowledge of the narrative, enabling the orchestration of events and experiences. This role is sometimes described as the "coach" or "trainer" in contexts involving training scenarios.
\end{itemize}

% 1) On a un manque de consensus
Despite the progress made to make interactive narratives better, the lack of consensus is still a notable
issue in this this field of research.

% 2) Les SNI ne sont pas assez comparés entre eux
Early work in interactive storytelling focused on the comparative analysis of INS \cite{magerkoComparativeAnalysisStory2007} and notes that these systems are often evaluated individually and are not rigorously compared to other approaches. This can be explained by the absence of an established benchmark (No Yardstick) \cite{koenitzChallengesIDNResearch2019}. One of the difficulties is that they use different stories. Since then comparisons work has been done, as between the ASD and PAST systems \cite{ramirezTellingInteractivePlayerspecific2012}, but the fact that PAST was implemented on top of ASD, and uses the same story, makes comparisons easier because it's based on the same representational basis. Despite the fact that the diversity of representation methods influences the difficulty of comparing these systems \cite{magerkoComparativeAnalysisStory2007}, there are studies comparing experience managers with heterogeneous designs like the one based on the "Generalized Experience Management" (GEM) framework \cite{moriStructuredAnalysisExperience2019, thueGeneralizedExperienceManagement2015}. \textcolor{black}{This structural analysis is based on the claim that all EM can be represented using a subset of the 5 blocks described by the framework and doesn't rely on the fact that the systems analyzed were explicitly designed with GEM in mind. After that, each block is analyzable and comparable after decomposition.}
Other comparisons are made by adopting a functional (or qualitative) approach rather than a structural one by examining properties \cite{luoReviewInteractiveNarrative2015, riedlInteractiveNarrativeIntelligent2013, robertsSurveyQualitativeAnalysis2008}. Our aim is to improve the INS comparison using a functional approach, based on a formalism capable of defining systems and their associated properties. \textcolor{black}{For a better understanding, structural analysis can be seen as a "white box", which focuses on the internal structure of the system in order to know the modules that make up the system and how they are developed, as does the GEM framework \cite{thueGeneralizedExperienceManagement2015}. Conversely, functional analysis can be seen as a "black box", because it doesn't have access to the structure of the system. Its goal is to analyze the system according to the specifications it must respect, defined by properties. In summary, structural analysis looks at how the system is built, while functional analysis looks at how the system behaves.}

All these approaches have the advantage of being modular and to combine related research. This allows research to progress by pooling the work of different researchers, as it was the case for the PAST system with the structural analysis from the GEM \cite{ramirezTellingInteractivePlayerspecific2012, thueGeneralizedExperienceManagement2015}. Still with the aim of comparison, GEM researchers are working on developing a complete evaluation plateform for EM (currently under development on github) \cite{moriEMGluePlatformDecoupling2022}. But even if a platform is finally released, it's still necessary for researchers in the field to adopt this tool and the associated representation which can be time-consuming, because the representation of the system must often be done in parallel with the design \cite{thueUnifiedUnderstandingExperience2018}. Furthermore, the systems that the researcher wishes to compare must also be converted and few of them are released after the publication of the linked scientific article \cite{moriEMGluePlatformDecoupling2022}. However, we assume that this effort is necessary, in order to coordinate and pool knowledge, enabling future research to be improved.

Our work is part of the effort to seek a common representation that can be used for all types of system design, or also for researchers wanting to coordinate their work and compare their results around a common framework. More specifically, we focus our efforts around a functional analysis methodology, by defining and comparing the system properties.

% 3) Pourquoi il y a autant de représentations ? et lesquelles ?
Current systems are designed for a specific domain \cite{thueGeneralizedExperienceManagement2015} in relation to their environment \cite{moriEMGluePlatformDecoupling2022}. This makes research fragmented, as each researcher creates their own design and the associated representation evolves in parallel with the chosen design. This diversity of representations \cite{thueUnifiedUnderstandingExperience2018} can be explained by the absence of a standard widely accepted by the community. We can find in the state of the art a multitude of existing story representations such as representations in the form of graphs \cite{riedlLinearStoryGeneration2006c}, planning problems \cite{porteousApplyingPlanningInteractive2010}, state machines \cite{TextWorldLearningEnvironment}, Markov Decision Processes \cite{robertsTargetingSpecificDistributions2006}, etc.

% 4) On doit avoir une représentation commune / quelle représentation reprendre ?
To improve the comparison of these different systems, a common representation approach is required. One project has proposed a strategy for comparing different EM representations \cite{thueUnifiedUnderstandingExperience2018} for this objective.

% TODO - quelle représentation prendre ???

% 5) En plus il y a une absence de terminologie
Another problem related to the lack of a commonly used approach is the lack of standardization of terminologies \cite{koenitzChallengesIDNResearch2019, thueGeneralizedExperienceManagement2015} which may cause misunderstandings in the work of the scientific community or redundancies in the justification of vocabulary choices (as is the case for researchers for example with the EM term \cite{thueUnifiedUnderstandingExperience2018, moriStructuredAnalysisExperience2019}).

% 6) Les recherches sont fragmentés et Faire le lien avec la section suivante
In this section, we pointed out the lack of consensus, the diversity of representations and the lack of standardized terminology that contribute to the fragmentation of research in INS. In the next section, we propose a model and its associated formalism, inspired by existing representations where we will explain our formalism choices. It gives us a common framework to represent INS, but also the basis of benchmarking methodology. According to this framework, we show in the \textit{Experiments} section how pertinent is the proposed framework based on state-of-the-art systems.

\section{Formalism}
\textcolor{black}{We propose here a model}, based on extended state machines. We define in this section the basis that would make future benchmarks of INS possible, through a functional analysis, based on the properties of these systems.

\subsection{System representation}

In order to describe INS, we propose a representation in the form of a state machine, by taking over certain parts of the other representations which we will develop later. We chose this representation because it benefits from the existing body of research studying this area, and has similarities with existing representations (in order to be able to integrate the work carried out) \textcolor{black}{such as the automated planning (this can be considered as the resolution of a search problem in a state transition system \cite{ModelInteractiveNarrative}) or Markov Decision Process (this is a stochastic model, where transitions between states, usually to represent the player's actions as in the GEM, are not necessarily deterministic).}

Let $\mathfrak{S} = (S, T , \gamma, s_{\text{init}}, S_{\text{goal}})$ be a state transition system where:

\begin{itemize}
    \item $S = \{s_1, s_2, \dots\}$ is a finite set of states.
    \item $T = A \cup E$ (with $A \cap E = \emptyset$) is a finite set of transitions:
    \begin{itemize}
        \item $A = \{a_1, a_2, \dots\}$ is a finite set of actions.
        \item $E = \{e_1, e_2, \dots\}$ is a finite set of events.
    \end{itemize}
    The EM can extend this set, such as $T' = T \cup T_{+EM}$ with $T'$ the new set and $T_{+EM}$, the transitions the manager wants to add. 
    \item $\gamma : S \times T \to S$ is a state transition function.
    The EM can update this function, such as $\gamma' : S \times T \to S$ becomes the new state transition function. A state transition is defined by a tuple $(s,t,s')$, where $s \in S$ is the current state, $t \in T$ the transition to be triggered and $s' \in S$ the target state.
    \item $s_{\text{init}} \in S$ is the initial state of the system.
    \item $S_{\text{goal}} \subseteq S$ is the set of final goal states of the system.
\end{itemize}

As mentioned earlier, this approach follows the principles of automatic planning where a planning problem is represented as a tuple $P = (D,i,g)$ where $D$ is a domain (represented by $S$ and $T$), $i$ is a set of literals indicating the initial conditions (represented by $s_{\text{init}}$) and $g$ is a set of literals indicating the objective (represented by $S_{\text{goal}}$).
We choose a disjoint perspective \cite{thueUnifiedUnderstandingExperience2018} in order to represent the EM and the system as two different entities. More specifically, we consider the EM as the system-side player and that's why we define two partitions in $T$ : Actions $A$ are associated with the player while events $E$ are dedicated to the EM. \textcolor{black}{We consider these two players as playing at the same time} (the player guided by his freedom, and the manager guided by the progression of the story), \textcolor{black}{but it's the EM who controls the system, whether to instantiate transitions (taking into account or not the player's choices), to update the transitions $T$ and the transition function $\gamma$}. The other advantage of this representation is that unlike other approaches that represent transitions only as player actions, we can, for example, represent a sequence of events where the player has no interaction to perform (depending on the the creator's narrative choices). \textcolor{black}{This representation also has disadvantages, such as the complexity of our system increasing exponentially depending on the size of the history, or that this representation doesn't take into account non-determinism. Moreover, this approach needs to obtain data to be used, whether by real players (can be challenging to obtain) or by simulations (only approximately representing the behavior of real players) as explained in the section \textit{Experiments}.} We add to this system different characteristics defined below.

\paragraph{Characteristic 1}
The system can only be in one state at a time.

\paragraph{Characteristic 2}
All states in $S$ are reachable from $s_{init}$ by at least one plan (see the \textit{Plan} section).

\paragraph{Characteristic 3}
We define $S_{\text{end}}$ as the set of states of the system $\mathfrak{S}$ that cannot lead to a new state. Formally, 
$
 S_{end} = \{ s \in S \mid \forall u \in T, \ \gamma(s, u) \text{ not defined} \}
$

Within this set $S_{\text{end}}$, we define two partitions:
\begin{itemize}
    \item Objective states $S_{\text{goal}} \subseteq S_{\text{end}}$ are the various states the player must reach.
    \item Problematic states $S_{\text{prob}} \subseteq S_{\text{end}}$ are states that block the progression of the narrative and make it impossible to reach an objective state. This can be refers as the "boundary problem" \cite{magerkoComparativeAnalysisStory2007}.
\end{itemize}

Formally, $S_{\text{end}} = S_{\text{goal}} \cup S_{\text{prob}}$ (with $S_{\text{goal}} \cap S_{\text{prob}} = \emptyset$)

\paragraph{Characteristic 4}
Unlike models where the states of the environment are partitioned into two sets (player states and EM states), where in player states, only actions $A$ can be triggered and in EM states, only events $E$ can be triggered, we decided to not use these partitions.
A state $s \in S \backslash S_{end}$, can have both outgoing action and event transitions.
The interest of this approach is to confront the choices/freedom of the player against the will of the creator and the manager (in some states, the manager must choose between triggering an action or an event). From the creator's point of view, this allows them more freedom since they are not forced to choose between two partitions of states during the creation process and to represent an absence of action by a transition. In the different representations, this "inaction" is presented as an action triggered ("no operation action" \cite{thueGeneralizedExperienceManagement2015}) or as a transition to switch to a new state ("donothing operator" \cite{ModelInteractiveNarrative}). The EM responsible for triggering transitions within the system, has to determine the strategy to implement based on the player's action (or inaction) received. A common strategy in games is to wait a certain amount of time for the player to act, otherwise it's considered as an inaction. \textcolor{black}{To formally define how the EM makes decisions, we introduce a policy function:}

\textcolor{black}{\qquad $\mu_{\text{EM}} : S \times (A \cup \{\varnothing\}) \to T_s$ where $T_s = \{ t \in T \mid \gamma(s, t) \text{ is defined} \}$}

\textcolor{black}{This function maps a current state $s \in S$ and a player-proposed action $a \in A \cup \{\varnothing\}$ (with $\varnothing$ denoting no action proposed, as an inaction) to a transition $t \in T_s$ that is enabled from state $s$. It ensures that only valid outgoing transitions can be selected, and allows the EM to validate a proposed player action, ignore it in favor of a narrative event, or react to player inactivity. The chosen transition $t \in T_s$ can then be activated via the transition function $\gamma$ to get a new state $s'$, such as $s' = \gamma(s, t)$.}

\subsection{Plan}
A plan is used to define the sequence of states that the system goes through during a player's experience. This definition of a plan, inspired by planning problem representation \cite{ModelInteractiveNarrative} is defined as $\pi$, a sequence $(s_0, s_1, \ldots, s_n)$ where $s_i \in S$ for $0 \leq i \leq n$. Each $s_i$ represents a state that is traversed successively by the system.

\textcolor{black}{\paragraph{Complete Plan} A plan $\pi$ is said to be complete if, for our system $\mathfrak{S}$, the initial state $s_0$ associated with the plan $\pi$ is equal to the initial state $s_{\text{init}} \in S$ (formally, $s_0 = s_{\text{init}}$), and the final state $s_n$ associated with the plan $\pi$ is equal to one of the objective states $S_{\text{goal}} \subseteq S$ (formally, $s_n \in S_{\text{goal}}$).}

\paragraph{Islands}
To illustrate the flexibility of our formalism and its applicability to various systems and concepts, we extend the definition of a plan, by adding the principle of islands described for the INS named ASD \cite{riedlDynamicExperienceManagement}. Islands are used to represent training objectives. They are "intermediate states in a search space, through which all solutions to a planning problem must pass." \cite{riedlDynamicExperienceManagement}. In our formalism, an island is represented as a subset of states in $S$. These islands help structure the search for solutions by introducing necessary intermediate steps before reaching the final objective. Formally:
\begin{itemize}
    \item $I = \{ I_1, I_2, \ldots, I_m \}$ where each $I_k \subseteq S$ represents an island, and $m$ is the total number of islands.
    \item Each state $s \in S$ belongs to at most one island: $\forall s \in S$, $\exists! I_k \in I$ such that $s \in I_k$.
    \item A state belonging to an island cannot be a final state ($s \in S_{\text{end}}$) or the initial state.
\end{itemize}
A sequence of objectives is an ordered list of islands $(I_1, I_2, \ldots, I_m)$ that the plan must traverse in this order before reaching the objective state. We can then add to our definition of a complete plan, in the case of an experiment with intermediate steps, the following point:
\begin{itemize}
    \item The passage through the intermediate steps is respected: \\
    Let $I = (I_1, I_2, \ldots, I_m)$ be a sequence of objectives and $\pi$ a plan represented by the sequence of states $(s_0, s_1, \ldots, s_n)$. The plan is considered complete if each island in $I$ appears at least once in the plan $\pi$, and the sequence of appearances of the islands in the plan is in ascending order.
\end{itemize}

Our proposed model provides a framework for formalizing INS, \textcolor{black}{with the objective that it can be reused to represent different systems}. This foundation enables the development of a benchmark methodology for evaluating and comparing different INS.

\section{Experiments}
% Pourquoi les simulations ? et pas des humains ?
In order to evaluate the performance of a system, we need to be able to launch interactive experiments. There are then two methods. The first way, the most obvious, is to have this system tested by human participants who will be able to interact with the system. Getting human feedback can be challenging. It often requires significant time and effort to get players to participate. In addition, it's possible that the EM or the creator wants to first obtain quick feedback on his interactive story to make modifications before publishing a final version. The second method consists in simulating experiments from probabilities on the player's potential actions that we must define beforehand, this allows to obtain quick feedback on the progress of the story, even if the results are only an approximation of real human users. We focus on the second option, as the results will help validate the formalism’s applicability and assess their relevance to the intended managers.

% Quelle histoire on utilise ? et quels SNI ?
To carry out our experiment, we chose the story of Little Red Riding Hood, which is common in the state of the art of interactive narrations \cite{riedlDynamicExperienceManagement,spierlingWorkshopPanelAuthoring2008,porteousAutomatedNarrativePlanning2021}, even if the researchers seem to use versions that are not entirely identical : we were therefore inspired by the version presented for ASD \cite{riedlDynamicExperienceManagement} to add the principle of the "islands"
\footnote{In this story, Little Red Riding Hood's goal is to give her grandmother a cake, but on her way she will meet the wolf. In this version, the wolf can eat Little Red Riding Hood, the grandmother, or both. The player plays the role of the hunter who must kill the wolf to save the person (or people) inside the wolf's belly. If the hunter kills the wolf before he meets Little Red Riding Hood then the story cannot continue, because the story loses its interest (the hunter kills the wolf before he commits any crime). The hunter must therefore kill the wolf after he devours someone (the island).}.
In addition, we implemented three managers simulating the behaviors described by researchers on their INS which are as follows:
\begin{itemize}
    \item \textbf{Vanilla} is an EM that doesn't apply any adaptation strategy or constraints to the player.
    \item \textbf{EM n°1} is inspired by ASD\cite{riedlDynamicExperienceManagement} which makes a fairy appear to resurrect the wolf, if the player kills it too early. It applies an adaptation strategy when the player is in a problematic state. If the player is stuck in a problematic state $s_1$, the system will apply an event transition with a probability equal to 1 to the previous state (denoted $s_0$). In addition, the system will apply a control to remove the transition from $s_0$ to $s_1$ \textcolor{black}{(denoted $a_{prob}$)} which previously allowed the player to find himself in a problematic state. \textcolor{black}{The manager can have full control over the system $\mathfrak{S}$ and can therefore change it dynamically, like update transitions $T$ and the transition function $\gamma$. In this case, when the system is in a problematic state, the manager extends the set $T$ with a new event transition $e_{fairy}$ such as $T := T \cup \{e_{fairy}\}$ and update the transition function to add the new event $e_{fairy}$ between two states, and remove the previous problematic transition, such as :}
    \textcolor{black}{
    \[
        \gamma' = \left\{
          \begin{array}{ll}
            s_0 & \text{if} \quad s = s_1 \quad \text{and} \quad t = e_{fairy} \\
            \text{not defined} & \text{if} \quad s = s_0 \quad \text{and} \quad t = a_{prob} \\
            \gamma(s,t) & \text{else}
          \end{array}
        \right.
    \]
    }

    \item \textbf{EM n°2} is inspired by Mimesis\cite{youngArchitectureIntegratingPlanbased2004a} which intervenes on the player's actions which could lead to a problematic state by cancelling their effects. It analyzes the effects of a player action before performing it. If an action $a_1$ has an arrival state of $s_1$ which is a problematic state, then the system will cause the action to fail. The effects of the action are not taken into account and the player remains in the same state. \textcolor{black}{Formally, the EM update the transition function such as :}
    \textcolor{black}{
    \[
        \gamma' = \left\{
          \begin{array}{ll}
            s & \text{if} \quad \gamma(s,t) \in S_{prob} \\
            \gamma(s,t) & \text{else}
          \end{array}
        \right.
    \]
    }
\end{itemize}

To conduct our experiments, we chose EM n°1 because it came from the article presenting the island principle with the associated Little Red Riding Hood story \cite{riedlDynamicExperienceManagement}. We also selected EM n°2 because of its popularity, and the fact that the article presenting it didn't use the same story \cite{youngArchitectureIntegratingPlanbased2004a} in order to show the interoperability of this formalism. Note that we didn't select a system that models the player. As a reference system, we implemented the Vanilla manager to determine a threshold that managers must exceed, otherwise they are considered as underperforming.
% Comment se déroule les simulations ?
To perform these simulations, we must define an associated probability for each transition. Since the objective of these experiments is to demonstrate the possibilities offered by this type of simulation, we arbitrarily define some probabilities. After adding the probabilities on the transitions of this system, we obtain a graph with for each node, the sum of the probabilities of the outgoing arcs is equal to 1. We then launch $n = 100$ simulations for each manager and collect the results. The results collected during a simulation are the player's plan (this plan is not necessarily complete) and for each instant (every step of the plan), we collect the probability matrix of the graph (that is to say the probabilities of all the transitions of the graph which can evolve over time according to the control exercised by the manager).

% Présentation des résultats

The simulations allow us to analyze the rate of them having a complete plan. Only the Vanilla manager has a success rate below 100\%. Additionally, we can see the number of adaptations made on average in each simulation for every EM (the number of times the EM has modified the probabilities of events). Only EM n°1 has a positive average, as it’s the only one offering this technique. The simulations also track state visit counts for each simulation. Thus we can analyze for the $n$ simulations of each EM, the visits carried out. We can then notice that the EM n°1 and n°2, unlike Vanilla, pass more than 100 times for $n = 100$ simulations on the initial state $s_0$. This is due to the fact that the strategies implemented by these EM allow to return to a previous state unlike Vanilla which cannot leave a problematic state. In addition, we note that EM n°2 never pass into a problematic state.

% Analyse des résultats
From these simulations, we can analyze the results obtained for each of these EM, as the success rates. EM n°1 and n°2 obtained a 100\% success rate unlike Vanilla. However, it's important to look at what brought these results. Indeed, both EM managed to handle boundary issues (caused by problematic states) but in different ways. EM n°2 has managed to avoid all problematic states, which makes this system the most robust \cite{luoReviewInteractiveNarrative2015}. However, this excessive robustness can translate into a lack of freedom for the player who sees their choices invalidated by the EM. As for it, EM n°1 is more resilient, allowing the user to make the choices they want without invalidating the effects of his action. In return, it exercises greater controllability \cite{luoReviewInteractiveNarrative2015} as proven by the number of adaptations made over the course of the story by altering the probabilities of events in order to manage the boundary issue problem.

% Bilan des résultats
The simulation results enable analysis of system properties, particularly EM robustness and controllability, and allow comparison across these properties. The properties are interconnected, in the case of the EM n°2, too much robustness leads to a limit of the user's freedom. This refers to the problem of narrative paradox \cite{aylettBeingThereParticipants2007}. This is the confrontation between the objectives desired by the EM (and the creator) with the freedom of the player, which can be an obstacle to the progression of the story. Future simulations will therefore focus on formalizing the properties of the different actors in this system, in order to interconnect the properties and their influences on others.

These experiments point out the pertinence and applicability of the proposed framework. For instance, we were able to describe the behaviors present in INS such as ASD and Mimesis in order to analyze the proposed repair strategies. This framework and methodology set up the basis to compare INS from different points of view and provide different scopes of analysis.

\section{Conclusion}
Interactive Narrative Systems offer a unique opportunity to engage users by allowing them to directly interact with dynamic narratives. However, the field suffers from a lack of standardization and a fragmentation of research efforts, limiting collective progress.

In this work, we proposed a methodology which aims to model and study the properties of INS, relying on state machine concepts. This framework not only allows to compare existing INS, but also to identify areas for improvement in terms of design and development.

Experiments conducted on the example of Little Red Riding Hood demonstrated the effectiveness of this formalization to analyze and evaluate different manager strategies. The results highlight the impact of design choices on the robustness of systems and the freedom of player actions, highlighting the importance of a balance between these properties.

This work provides a basis for the development of common methodologies in the field of INS. In the future, further efforts will be needed to integrate properties from the creator, player and manager perspectives to improve the analysis and to extend this formalization to even more complex scenarios. By bringing the community together around shared standards, we hope to unify research and enrich the possibilities offered by interactive narratives, especially with the emergence of LLM.

\section{Acknowledgments}
This work was supported by funding from the ANRT (Association Nationale de la Recherche et de la Technologie) as part of my doctoral research. I would like to express my deepest gratitude to Dr. Mark Riedl, Professor at the Georgia Tech School of Interactive Computing, for granting me access to his work, which significantly informed and enriched the formalism developed in this study.

\printbibliography

\end{document}